\documentclass[pra,preprint]{revtex4} 
 
\usepackage{amsmath} 
\usepackage{graphicx} 

\begin{document} 

\title{Characterization of pure quantum states of multiple qubits \\
using the Groverian entanglement measure} 
 
\author{Yishai Shimoni, Daniel Shapira and Ofer Biham} 
\affiliation{Racah Institute of 
Physics, The Hebrew University, Jerusalem 91904, Israel} 
 
\begin{abstract} 
The Groverian entanglement measure, $G(\psi)$, is applied to 
characterize a variety of pure quantum states 
$| \psi \rangle$
of multiple qubits.
The Groverian measure is calculated analytically 
for certain states of high symmetry, 
while for arbitrary states 
it is evaluated using a numerical procedure.
In particular, it is calculated
for the class of 
Greenberger-Horne-Zeilinger states, the W states
as well as for random pure states of $n$ qubits. 
The entanglement generated by Grover's algorithm is
evaluated by calculating $G(\psi)$ for
the intermediate states that are obtained after
$t$ Grover iterations, for various initial states
and for different sets of the marked states.
\end{abstract} 
 
\pacs{03.67.Lx, 89.70.+c} 
 
\maketitle 
 
\section{Introduction} 

In the past decade it was demonstrated that for certain computational tasks,
quantum algorithms are more efficient than classical algorithms.
Two examples of this effect are provided by Shor's factorization
algorithm
\cite{Shor1995}
and Grover's search algorithm
\cite{Grover1996,Grover1997a}.
While the understanding of quantum algorithms is still incomplete,
there are indications that quantum entanglement plays a crucial
role in making them powerful.
Quantum algorithms generate entangled states that
involve large numbers of qubits.
To assess the role of entanglement in the algorithms it would
be useful to develop ways to quantify it,
which are based on operational considerations.
Such considerations in the context of quantum communication
between two parties
were successfully applied to develop measures of entanglement
for bi-partite systems
\cite{Bennett1996a,Bennett1996b,Popescu1997}.
It is thus expected that operational considerations in the context
of quantum computation, that involves multiple qubits,
may lead to useful entanglement measures for multi-partite 
systems.
Recent work based on axiomatic considerations has provided a set
of properties that entanglement measures should satisfy
\cite{Vedral1997,Vedral1998,Vidal2000,Horodecki2000}.
These properties include the requirement that any entanglement
measure should vanish for product states, 
it should be invariant under local unitary operations
and should not increase
as a result of any sequence of local operations complemented by only classical 
communication between the parties.
These properties provide useful guidelines in the search for
operational measures of entanglement based on quantum algorithms.
Measures that satisfy the properties specified above are called
entanglement monotones.

Recently, it was shown that the entanglement of a pure quantum
state 
$| \psi \rangle$
of $n$ qubits can be quantified by the utility of 
this state as an initial state for Grover's search algorithm
\cite{Biham2002}.
Suppose that a given state
$| \psi \rangle$
is used as the initial state.
However, before starting the search, 
one is allowed to perform local unitary operations on
each qubit in order to maximize the
probability,
$P_{\rm max}$,
of having a successful quantum search.
It was shown that 
if 
$| \psi \rangle$
can by transformed to
some other state
$| \phi \rangle$
by local operations and classical communication
then
$P_{\rm max}(\psi) \le P_{\rm max}(\phi)$.
This observation was used in order to construct
the Groverian entanglement measure $G(\psi)$
of pure states. 
The Groverian measure is an 
entanglement monotone, and is equivalent
to an entanglement measure proposed previously
in Refs.
\cite{Vedral1997,Vedral1998},
which is based on distance measures between 
$| \psi \rangle$
and the nearest disentangled state.

In this paper we present an explicit representation of 
$G(\psi)$ and calculate it for a variety of pure entangled
states using a combination of analytical and numerical methods.
The difficulty is that the calculation of the Groverian measure
is based on the maximization of
a multi-variable function.
The number of variables over which the maximization is taken
increases with the number of qubits, $n$, in the state 
$| \psi \rangle$.
The measure $G(\psi)$ is calculated analytically for various
quantum states that exhibit high symmetry,
such as the
Greenberger-Horne-Zeilinger state
\cite{Greenberger1990} 
and the W state
\cite{Zeilinger1997,Dur2000}
of $n$ qubits.
The numerical procedure is applied to the calculation of
$G(\psi)$ for the states that are produced by Grover's
algorithm as a function of the number of iterations, $t$,
when the initial state is either a product state or
an entangled state
and for different sets of marked states.
Finally, random states of $n$ qubits are also examined.
It is found that homogeneous random sampling of
pure states of $n$ qubits produces highly entangled states
for which
$G(\psi)$ approaches 1 for large $n$.

The paper is organized as follows. 
The quantum search algorithm, using arbitrary initial states
is described in Sec. II. The Groverian entanglement measure
is presented in Sec. III.
An explicit representation of the Groverian measure is
introduced in Sec. IV. This explicit representation is
used in Sec. V to perform analytical calculations of 
the Groverian measure for certain pure states
of high symmetry. Numerical calculations of
$G(\psi)$ for arbitrary pure states are presented in Sec. VI.
The results are summarized in Sec. VII.

\section{Quantum search using an arbitrary initial state} 

Consider a search space $D$ containing $N$ elements.  We assume, for 
convenience, that $N = 2^n$, where $n$ is an integer. In this way, we 
may represent the elements of $D$ using an $n$-qubit {\em register} 
containing the indices, $i=0,\dots,N-1$.  We assume that a subset of 
$r$ elements in the search space are marked, that is, they are 
solutions to the search problem.  The distinction between the marked 
and unmarked elements can be expressed by a suitable function, 
$f: D \rightarrow \{0,1\}$, 
such that $f=1$ for the marked elements, and $f=0$ for the rest. 
The search for a marked element now becomes a search for an element 
for which $f=1$.  To solve this 
problem on a classical computer one needs to evaluate $f$ for each 
element, one by one, until a marked state is found.  Thus, on average, 
$N/2$ evaluations of $f$ are required and $N$ in the worst case.
For a quantum computer, on which
$f$ to be evaluated 
\emph{coherently}, 
it was shown that a sequence of unitary operations 
called Grover's algorithm
can locate a marked element using only $O(\sqrt{N/r})$ coherent 
queries of $f$ \cite{Grover1996,Grover1997a}.  
 
To describe the operation of the quantum search algorithm we first 
introduce a register, 
$\left| i \right\rangle = 
\left| i_{1} \ldots   i_{n} \right\rangle$, 
of $n$ qubits, 
and an 
\emph{ancilla} qubit, 
$|q\rangle$, 
to be used in the computation.  
We also introduce a 
\emph{quantum oracle}, 
a unitary operator $O$ which functions as a 
black box with the ability to 
\emph{recognize} solutions to the search 
problem.  
The oracle performs the following 
unitary operation on computational basis states of the register, 
$\left| i \right\rangle$, 
and the ancilla, 
$\left| q \right\rangle$:
 
\begin{equation}
O \left| i \right\rangle \left| q \right\rangle = 
\left| i \right\rangle   \left| q \oplus f(i) \right\rangle, 
\label{eq:bborac} 
\end{equation} 

\noindent
where $\oplus$ denotes addition modulo 2. 
The oracle recognizes marked states in the sense that if 
$| i \rangle$ 
is a marked element of the search space, 
namely $f(i) = 1$, 
the oracle flips the ancilla qubit from 
$\left| 0 \right\rangle$ 
to 
$\left| 1 \right\rangle$ 
and vice versa, 
while for unmarked states the ancilla is unchanged.  
The ancilla qubit is initially 
set to the state 

\begin{equation} 
| - \rangle_q = { \frac{1}{\sqrt{2}} } (\left| 0 \right> - \left| 1 \right>).  
\end{equation} 

\noindent
With this choice, the action of the oracle is: 

\begin{equation} 
O |i\rangle |-\rangle_q 
= (-1)^{f(i)} |i\rangle |-\rangle_q.
\end{equation} 

\noindent
Thus, the only effect of the oracle is to apply a phase of $-1$ if 
$|i\rangle$ 
is a marked basis state, and no phase change if 
$|i\rangle$ 
is unmarked.  
Since the state of the ancilla does not change, 
one my omit it and write the action of the oracle as 
$O|i\rangle = (-1)^{f(i)}|i\rangle$.  

The original algorithm, as introduced by Grover, includes an 
initialization stage in which the
$n+1$ qubits of the register and the ancilla are prepared in the
state $|0\rangle^{\otimes n}|0\rangle_q$.
Then, the following procedure is performed: 
a Hadamard gate    
$H = \frac{1}{\sqrt2}
\left(
\begin{smallmatrix} 
1 & 1 \\ 
1 & -1 
\end{smallmatrix}
\right)$ 
is applied on each qubit in the register, 
and the gate $HX$ on the ancilla, 
where 
$X=
\left(
\begin{smallmatrix} 
0 & 1 \\ 
1 & 0 
\end{smallmatrix}
\right)$ 
is the {\sc not} gate.  
The matrices are expressed in the computational basis 
($|0\rangle,|1\rangle$).  
The resulting state is: 

\begin{equation}
|\eta \rangle |-\rangle_q, 
\label{eq:etam} 
\end{equation} 

\noindent
where

\begin{equation}
| \eta \rangle = \frac{1}{\sqrt{N}} \sum_{i=0}^{N-1} |i\rangle. 
\label{eq:eta} 
\end{equation} 

\noindent
The state $\eta$ is considered as the intial state of the algorithm.

Here we consider a generalized algorithm in which an arbitrary,
possibly entangled state

\begin{equation}
| \psi \rangle = \sum_{i=0}^{N-1} a_i |i\rangle,
\label{eq:phi}
\end{equation}

\noindent
is used as the initial state instead of 
$| \eta \rangle$.
The ancilla is still prepared as before, namely its state is
$|-\rangle_q$. 
The algorithm itself consists of $\tau$ repetitions of the
following Grover iteration:

\begin{enumerate} 
\item \label{en:rot1} Apply the oracle, which has the effect of 
  rotating the marked states by a phase of $\pi$ radians.  Since the 
  ancilla is always in the state $|-\rangle_q$ the effect 
  of this operation 
  may be described by the following unitary operator 

\begin{equation}
I_f^{\pi} = \sum_{i=0}^{N-1} (-1)^{f(i)} | {i} \rangle \langle {i} |, 
\end{equation} 

acting only on the register. 

\item Carry out the following steps:  (i) apply the 
  Hadamard gate to each qubit in the register; (ii) Rotate the 
  $\left| 00 \ldots 0 \right\rangle$ state  
  of the register by a phase of $\pi$ 
  radians.  This rotation is similar to step 1, except for the fact that 
  here it is performed on a known state. It takes the form
 
\begin{equation}
I_{0}^{\pi} = - 2 |0\rangle \langle 0| + \sum_{i=0}^{N-1} |i\rangle \langle i|, 
\end{equation} 

where the second term on the right hand side is the identity operator, denoted
by $I$. 
(iii) Apply the Hadamard gate again on each qubit in the register. 

The resulting operation is

\begin{equation}
- H^{\otimes n} I_{0}^{\pi} H^{\otimes n} = 
    -I + 2 H^{\otimes n} |0\rangle \langle 0| H^{\otimes n} =
    -I + 2 | \eta \rangle \langle \eta |.
\end{equation}

When this operator is applied on the state 
$| \psi \rangle$
it results in the state
 
\begin{equation}
- H^{\otimes n} I_{0}^{\pi} H^{\otimes n} | \psi \rangle = 
\sum_{i=0}^{N-1} (2 \bar{a} - a_i) | i \rangle,
\end{equation}

\noindent
where 

\begin{equation}
\bar a = {\frac{1}{N}} \sum_{i=0}^{N-1} a_i. 
\label{eq:abar}
\end{equation}

Thus, each amplitude is rotated by $\pi$ around the
average of all amplitides of the quantum state.
\end{enumerate} 

The combined operation on the register
in one Grover iteration 
is described by 

\begin{equation}
U_G = - H^{\otimes n} I_{0}^{\pi} H^{\otimes n} I_f^\pi. 
\label{eq:U_G}
\end{equation}

\noindent
After the completion of $\tau$ Grover iterations,
the register is measured in the computational basis. 
The optimal number of iterations is
\cite{Grover1997a,Boyer1998,Zalka1999}

\begin{equation}
\tau = \left\lfloor
\left(\frac{\pi}{2}-\sqrt{\frac{r}{N-r}}\right)/ \cos^{-1} (1-2r/N)
\right\rfloor, 
\label{eq:optitexact} 
\end{equation} 

\noindent
or, approximately for $r \ll N$

\begin{equation}
\tau = \left\lfloor
\frac{\pi}{4} \sqrt{\frac{N}{r}}
\right\rfloor, 
\label{eq:optit} 
\end{equation} 

\noindent
where $\lfloor x \rfloor$
is the largest integer which is smaller than $x$.
Using the original initialization process,
at this optimal time, a marked state
can be found with almfost certainty, or more 
precisely with probability

\begin{equation}
P_{\rm s}(\eta) = 1 - O \left({ \frac{1}{\sqrt{N}} } \right). 
\end{equation}

\noindent
With this performance,
Grover's algorithm was found to be 
optimal
\cite{Zalka1999},
in the sense that it is as efficient as 
theoretically possible
\cite{Bennett1997}. 
A variety of applications were developed, in which
the algorithm is used in the solution of other problems
\cite{Grover1997b,Terhal1998,Brassard1998,Grover2000,Cerf2000}.
The algorithm was also generalized by allowing an arbitrary
(but fixed) unitary transformation to take place of the Hadamard
transform and an arbitrary phase rotation instead of the $\pi$
inversion
\cite{Grover1998,Long1999,Gingrich2000,Biham2001}.

When a general pure state
$| \psi \rangle$
is taken as the initial state for Grover's algorithm
the success probability is reduced. 
In this case,
the probability $P(t)$ to find a marked state if
a measurement is taken after $t$ iterations 
depends
not only on the initial state
$| \psi \rangle$ 
and the number of marked states,
but also on the specific identity of the set of marked states.
For a given choice of the set of maked states, 
the states 
$| \psi(t) \rangle$ 
obtained after $t$ Grover iterations, starting 
with an arbitrary initial state
$| \psi(0) \rangle$, 
were calculated using recursion equations 
\cite{Biham1999}.
The optimal time to measure as well as the
maximal probability of success were found to depend
both on the initial state 
$| \psi \rangle$
and on the specific choice of the set of marked states.
Of course, 
in a real search, the set of marked states is
unknown,
although it is assumed that the number of marked states is known. 
To evaluate the success probability and the
optimal time to measure one needs to
perform an average over all possible choices of the
set of marked states. 
Under these conditions, 
the optimal time to measure was found to be equal to
$\tau$,
namely, after the same number of iterations as for the case in which
the initial state is $| \eta \rangle$  
\cite{Biham2003}.
However,  
the success probability
$P_{\rm s}(\psi)$ 
is reduced.
It can be expressed in terms of the amplitudes
of the initial state
according to

\begin{equation}
P_{\rm s}(\psi) = N |\bar{a}|^2 + O \left({ \frac{1}{\sqrt{N}} } \right).
\end{equation}

\noindent
Thus, the success probability of the search depends only on the 
first moment of the distribution of the amplitudes and not on
higher moments. 
Moreover, it does not depend on the number of marked states,
up to a correction of order
$r/N$
\cite{Biham2003}.

\section{The Groverian entanglement measure}
\label{sec:groverian} 

Consider $n$ parties sharing a pure quantum state 
$|\psi\rangle$ of $n$ qubits, where 
each party is in possession of one qubit.  
The parties use 
\emph{those particular $n$ qubits} 
to perform a quantum search in the space of $N=2^n$ 
elements.  
Prior to the search, each party may perform local unitary 
operations on the qubit in their possession.  After they complete the 
local processing of their qubits, all parties send (or teleport) their 
qubits to the search processing unit.  
The only processing available 
in this unit is Grover's search iterations and the subsequent 
measurement. Thus, the only way the qubits are allowed to interact is 
through Grover iterations. 
 
The local pre-processing can be expressed by

\begin{equation}
V=U_1 \otimes U_2 \otimes \cdots \otimes U_n, 
\label{eq:V}
\end{equation}

\noindent
where $U_k$ is an arbitrary local unitary gate acting 
on the $k$th qubit.  
The initial state inserted into the Grover iterations
is then
$V | \psi \rangle$.
These local operations are chosen such that the success
probability
$P_{\rm s}$
of the algorithm will be maximized. 
The maximal success probability under these
conditions will be

\begin{equation}
P_{\rm max}(\psi) = \max_{U_1,\dots,U_n} 
P_{\rm s}(U_1 \otimes U_2 \otimes \cdots \otimes U_n|\psi\rangle).
\end{equation}

\noindent
It turns out that 
$P_{\text{max}}(\psi)$ 
can be used to quantify
the entanglement present in the state
$|\psi\rangle$. 
To make this assertion more precise, let us write $P_{\max}(\psi)$ in terms 
of the operator $U_G^{\tau}$ representing $\tau$ Grover iterations.  
For simplicity consider the case in which there is only a single
marked state
$|m\rangle$.
Since the marked state is unknown, the probability
$P_{\text{max}}(\psi)$ 
should be averaged over all possible identities of $m$,
namely
$m=0,1,\dots,N-1$.
Performing this average we obtain that 

\begin{equation}
P_{\max}(\psi) = \max_{U_1,\ldots,U_n} \frac{1}{N} \sum_{m=0}^{N-1} 
\left| \langle m| U_G^{\tau} (U_1 \otimes U_2 \otimes \cdots \otimes 
U_n) |\psi\rangle \right|^2, 
\label{eq:Pmaxdef} 
\end{equation} 

\noindent
where the maximization is over all local unitary operations $U_1, 
\ldots, U_n$ on the respective qubits. 
In order to evaluate $P_{\rm max}(\psi)$ we recall that

\begin{equation}\label{eq:getms} 
U_G^{\tau} \left| \eta \right> = |m\rangle + 
O\left(\frac{1}{\sqrt{N}}\right), 
\end{equation} 

\noindent
where the second term is a small correction. 
Multiplying this equation 
by $(U_G^{\tau})^{\dagger}$ 
and taking the Hermitian conjugate gives 

\begin{equation} 
\langle m| U_G^{\tau} 
= \langle\eta| + O\left(\frac{1}{\sqrt{N}}\right). 
\end{equation} 

\noindent
Substituting into Eq.~(\ref{eq:Pmaxdef}) gives, for a general state 
$|\psi\rangle$, 

\begin{equation}
P_{\max}(\psi) = \max_{U_1,\ldots,U_n} \frac{1}{N} \sum_{s=0}^{N-1} 
\left| \langle \eta | U_1 \otimes U_2 \otimes \cdots \otimes U_n 
|\psi\rangle \right|^2 + O\left(\frac{1}{\sqrt{N}}\right). 
\label{eq:pmax2} 
\end{equation} 

\noindent
However, 
$|\eta\rangle$ 
is a product state, so that 
$U_1^\dag \otimes U_2^\dag \otimes \cdots \otimes U_n^\dag|\eta\rangle$ 
is another product state. 
Therefore, the maximization in 
Eq.~(\ref{eq:pmax2}) may, equivalently, be expressed by

\begin{equation}
P_{\max}(\psi) = \max_{|e_1, \ldots, e_n\rangle} \left| \langle 
e_1,\ldots, e_n|\psi \rangle \right|^2 + 
O\left(\frac{1}{\sqrt{N}}\right), 
\label{eq:pmax3} 
\end{equation} 

\noindent
where the maximization now runs over all product states, 
$|e_1,\ldots,e_n\rangle=|e_1\rangle\otimes\cdots\otimes|e_n\rangle$, 
of the $n$ qubits.  

In Ref.
\cite{Biham2002} 
it was shown that the
maximum success probability, 
$P_{\text{max}}(\psi)$, 
can be used to define an entanglement measure, 
the {\em   Groverian entanglement}, 
for arbitrary pure multiple qubit states. 
The Groverian entanglement of a 
state $|\psi\rangle$ is given by: 

\begin{equation}
G (\psi) \equiv \sqrt{1 - P_{\max}(\psi)}. 
\label{eq:Gofphi}
\end{equation} 

\noindent
Since $P_{\max}(\psi)$ takes values in the range 
$0 \le P_{\max}(\psi) \le 1$, it follows that 
$0 \le G (\psi) \le 1$.  
It is clear
from the definition that 
all the states 
$| \psi \rangle$
that can be reached from
$| \psi \rangle$
by local unitary operations
share the same measure, given
by
$G (\psi)$.
It is also easy to see that  
for all
product states
$G (\psi) =0$. 
In Ref. 
\cite{Biham2002}
it was shown that the Groverian entanglement measure is closely related
to an entanglement measure introduced previously in Refs. 
\cite{Vedral1997,Vedral1998}. 
This relation was used in order to 
demonstrate that 
$G(\psi)$
is an entanglement monotone,
namely it cannot be increased by local operations and classical 
communication. 
Therefore, the Groverian measure is a good entanglement measure
for pure quantum states of multiple qubits. 

\section{Explicit representation of the Groverian measure}

Consider a pure state

\begin{equation}
|\psi\rangle = \sum_{i=0}^{N-1} a_i |i\rangle 
\end{equation}

\noindent
of $n$ qubits,
where

\begin{equation}
|i\rangle = |i_1,i_2,\dots,i_n\rangle
\end{equation}

\noindent
and $i_k$ is the $k$th most significant bit 
of the binary integer $i$.
The Groverian entanglement measure 
$G(\psi)$ 
is given 
by Eq.
(\ref{eq:Gofphi}),
where
$P_{\rm max}(\psi)$
is given 
by Eq.
(\ref{eq:pmax3}).
To obtain an explicit formula for 
$G(\psi)$ 
we will consider
a general product state of $n$ qubits

\begin{equation}
|e\rangle = |e_1\rangle \otimes \dots \otimes|e_n\rangle. 
\label{eq:productstate}
\end{equation}

\noindent
The single qubit states can be represented by

\begin{equation}
|e_k\rangle = 
\cos{\theta_k}|0\rangle_k
            + e^{i\varphi_k}\sin{\theta_k}|1\rangle_k, 
\label{eq:sbs1}
\end{equation}

\noindent
where $k=1,2,\dots,n$, 
and global phases are ignored.
Note that our angle
$0 \le \theta_k \le \pi/2$
is a half of the angle $\theta$
used in the Bloch sphere representation,
while $\varphi_k$
is in the range
$0 \le \varphi_k \le 2 \pi$.

The product state takes the form

\begin{eqnarray}
|e_1\rangle \otimes \dots \otimes|e_n\rangle &=& 
\cos{\theta_1}\dots\cos{\theta_n}|0\ldots 0\rangle +\\ \nonumber 
&+&\cos{\theta_1}\dots e^{i\varphi_n}\sin{\theta_n}|0\ldots 01\rangle \nonumber \\
&+& \dots +\\ \nonumber
&+&e^{i\varphi_1}\sin{\theta_1}\dots e^{i\varphi_n}\sin{\theta_n}|1\ldots 1\rangle.
\label{eq:sbs2}
\end{eqnarray}

\noindent
Therefore, the product state 
$|e\rangle$
can by written in the form

\begin{equation}
|e\rangle = \sum_{i=0}^{N-1} c_i |i\rangle, 
\end{equation}

\noindent
where the coefficient of the
basis state
$|i\rangle = |i_1,i_2,\dots,i_n\rangle$
is

\begin{equation}
c_i = \prod_{k=1}^n (\cos \theta_k)^{\bar i_k} (e^{i \varphi_k} \sin \theta_k)^{i_k}, 
\label{eq:c_i}
\end{equation}

\noindent
where
$\bar i_k = 1 - i_k$.
The overlap between the given state 
$| \psi \rangle$
and a product state
$| e \rangle$
is defined by

\begin{equation}
\langle e| \psi \rangle = \sum_{i=0}^{N-1} a_i \ c_i^{\ast},
\end{equation}

\noindent
where $c_i^{\ast}$
is the complex conjugate of $c_i$.
The calculation of the Groverian measure involves the the 
maximization of the function

\begin{equation}
P(\theta_1,\dots,\theta_n,\varphi_1,\dots,\varphi_n,\psi) =
|\langle e_1, \ldots, e_n | \psi\rangle|^2 
\end{equation}

\noindent
with respect to the variables $\theta_k, \varphi_k$, 
$k=1,\dots,n$.
Note that for $\theta_k=\pi/4$ and $\varphi_k=0$, $k=1,\dots,n$,
this function coincides with the success probability of Grover's 
algorithm starting with the initial state $| \psi \rangle$,
given by
$P_{\rm s}(\psi)= |\langle \eta|\psi \rangle|^2$.
The maximal probability of success can now be written as

\begin{equation}
P_{\max}(\psi) = 
\max_{\theta_1, \ldots, \theta_n, \varphi_1, \ldots, \varphi_n } 
P(\theta_1,\dots,\theta_n,\varphi_1,\dots,\varphi_n,\psi), 
\label{eq:pmax} 
\end{equation} 

\noindent
up to a correction term of order
$1/\sqrt{N}$,
and the maximization is taken in the range 
$0 \le \theta_k \le \pi/2$
and
$0 \le \varphi_k < 2 \pi$.

The case of states 
$| \psi \rangle$ 
in which all the amplitudes 
$a_i$ are real is simpler.
For such states, the
product state 
$|e_1,\dots,e_n\rangle$
for which the maximum in Eq.
(\ref{eq:pmax})
is obtained has real amplitueds as well, 
namely all the angles
$\varphi_k=0$ or $\pi$. 
Therefore, in this case the maximization over
$\varphi_k$
is reduced to a discrete maximization over
the binary choice of
$\exp(i\varphi_k)=\pm 1$.
This term can be removed by doubling the range
of $\theta_k$ to
$-\pi/2 \le \theta_k \le \pi/2$,
thus allowing 
$\sin \theta_k$ 
to be
both positive and negative,
for the same value of
$\cos \theta_k$.
Thus, for states
$| \psi \rangle$ 
in which all the amplitudes
$a_i$
are real

\begin{equation}
P_{\max}(\psi) = 
\max_{\theta_1, \ldots, \theta_n} 
P(\theta_1,\dots,\theta_n,\psi), 
\label{eq:pmaxreal} 
\end{equation} 

\noindent
where the maximization is over the range 
$-\pi/2 \le \theta_k \le \pi/2$
and 
$\varphi_k=0$, $k=1,\dots,n$.

\section{Analytical calculations of the Groverian measure}

Having found an explicit expression for the 
Groverian entanglement measure, we can now 
use it in order to characterize certain quantum states 
that are encountered in various contexts of 
quantum computation and communication. 

\subsection{Two-qubit states}

Consider a
state $| \psi \rangle$ of two qubits
in which all the amplitudes are real. 
According to 
Eq.~(\ref{eq:pmaxreal}) 
we can write the overlap
of $| \psi \rangle$ 
with a tensor product of two single qubit states as

\begin{equation}
P(\theta_1,\theta_2,\psi) = 
(a_{00}\cos\theta_1\cos\theta_2+
a_{01}\cos\theta_1\sin\theta_2+
a_{10}\sin\theta_1\cos\theta_2+
a_{11}\sin\theta_1\sin\theta_2)^2.
\label{eq:2bit}
\end{equation}

\noindent
Using standard trigonometric identities we obtain

\begin{equation}
P(\theta_p,\theta_m,\psi) = 
  \left( \frac{a_{00}-a_{11}}{2}\cos \theta_p 
+ \frac{a_{00}+a_{11}}{2}\cos \theta_m 
+ \frac{a_{10}+a_{01}}{2}\sin \theta_p 
+ \frac{a_{10}-a_{01}}{2}\sin \theta_m \right)^2,
\label{eq:Pofthetas}
\end{equation}

\noindent
where 

\begin{eqnarray}
\theta_p &=& \theta_1+\theta_2 \nonumber \\ 
\theta_m  &=& \theta_1-\theta_2. 
\end{eqnarray}

\noindent
To obtain 
$P_{\rm max}(\psi)$
one needs to 
maximize 
$P(\theta_1,\theta_2,\psi)$ 
with respect to $\theta_1$ and $\theta_2$,
or equivalently, 
with respect to $\theta_p$ and $\theta_m$. 
This maximization is done by solving
 
\begin{eqnarray}
\frac{\partial P(\theta_p,\theta_m,\psi)}
{\partial\theta_p} &=& 0 \nonumber \\ 
\frac{\partial P(\theta_p,\theta_m,\psi)}
{\partial\theta_m} &=& 0.
\end{eqnarray}

\noindent
The maximun is found for 
$\theta_p$
and
$\theta_m$
that satisfy

\begin{eqnarray}
\cos\theta_p &=& \frac{a_{00}-a_{11}}{\sqrt{(a_{10}+a_{01})^2+(a_{00}-a_{11})^2}} \nonumber \\ 
\cos\theta_m  &=& \frac{a_{00}+a_{11}}{\sqrt{(a_{10}-a_{01})^2+(a_{00}+a_{11})^2}}.
\end{eqnarray}

\noindent 
Inserting these values into
Eq.
(\ref{eq:Pofthetas})
gives rise to

\begin{equation}
P_{\rm max}(\psi) = \frac{1}{2}\left(1 + \sqrt{1 - 4 |\det D|^2}\right),
\label{eq:2bitmax}
\end{equation}

\noindent
where the matrix $D$ takes is given by 

\begin{eqnarray} 
D = \left( 
\begin{array}{ll} 
a_{00}  & a_{01} \\ 
a_{10}  & a_{11}                          
\end{array} 
\right). 
\end{eqnarray} 

\noindent
Eq. (\ref{eq:2bitmax})
is valid also in case that the amplitudes are complex,
however in this case the direct maximization is tedius.
This result can be shown using the Schmidt decomposition
\cite{Nielsen2000}
and the fact that
$P_{\rm max}(\psi)$
is equal to the square of the maximal
Schmidt coefficient
\cite{Barnum1999,Vidal2000b}.
Therefore, the von Neumann entropy, $S$, of the reduced density matrix
can be expressed by
\cite{Biham2002}

\begin{equation}
S = -P_{\rm max} \ln_2 P_{\rm max} - 
     (1- P_{\rm max}) \ln_2 (1-P_{\rm max}). 
\end{equation}

Consider a generalized Bell state of the form 

\begin{equation}
|\psi\rangle = a_{00} |00> + a_{11} |11>. 
\end{equation}

\noindent
Inserting these amplitudes into 
Eq.~(\ref{eq:2bitmax})
we obtain that

\begin{equation}
P_{\rm max}(\psi) = \max(|a_{00}|^2,|a_{11}|^2).
\label{eq:bell}
\end{equation}

\noindent
Therefore, the Bell states 
$|\phi_{\pm}\rangle$,
for which
$|a_{00}|=|a_{11}|=1/\sqrt{2}$,
as well as the two other Bell states,
are characterized by
$P_{\rm max} = 1/\sqrt{2}$.

In case that $|\psi \rangle$ is a product state, its amplitudes
can be expressed by
$a_{ij}=b_i c_j$, $i,j=0,1$,
where 
$b_i$
and $c_j$
are the amplitudes of the two
single qubit states that form the state 
$| \psi \rangle$.
Plugging this product into
Eq.
(\ref{eq:2bitmax})
it is easy to see that for product states
$P_{\rm max} = 1$.

\subsection{Multiple-qubit states}

\subsubsection{Generalized GHZ states:}

The GHZ state of $n$ qubits 
takes the form

\begin{equation}
|\psi_{\rm GHZ}\rangle 
= {\frac{1}{\sqrt{2}}}{(|0\ldots0\rangle + |1\ldots1\rangle)}. 
\label{eq:ghzDef}
\end{equation}

\noindent
This is a generalization of the Bell state $|\phi_{+}\rangle$
to systems of more than two entangled qubits. These states
can be further generalized to a continuous class of states
of the form

\begin{equation}
|\psi\rangle = a_0 |0\ldots0\rangle + a_{N-1} |1\ldots1\rangle, 
\label{eq:Genghz}
\end{equation}

\noindent
where $|a_0|^2+|a_{N-1}|^2=1$. 
In this case the overlap function takes the form

\begin{equation}
P(\theta_1,\dots,\theta_n,\varphi_1,\dots,\varphi_n,\psi) = 
\left(a_0 \prod_{k=1}^{n} \cos{\theta_k}
+ a_{N-1} \prod_{k=1}^{n} e^{i \varphi_k} \sin{\theta_k}   \right)^2.
\end{equation}

\noindent
To obtain $P_{\rm max}(\psi)$
we solve the equations
${\partial P(\theta_1,\dots,\theta_n,\psi)}/{\partial\theta_k}=0$
and
${\partial P(\theta_1,\dots,\theta_n,\psi)}/{\partial\varphi_k}=0$
for 
$k=1,\dots,n$.
The solution is 
$\theta_k=0 (\pi/2)$, $k=1,\dots,n$,
when
$|a_0|$
is larger (smaller)
than
$|a_{N-1}|$.
Therefore,
$P_{\rm max}(\psi) = \max \{ |a_0|^2,|a_{N-1}|^2 \}$
and 
$G(\psi)$
can be obtained from Eq.
(\ref{eq:Gofphi}).
Its maximal value is obtained when
$|a_0|=|a_{N-1}|=1/\sqrt{2}$, where
$G(\psi_{\rm GHZ}) = 1/\sqrt{2}$,
namely independent of the number of qubits.

\subsubsection{The W states:}

The W state of $n$ qubits is the symmetrical 
superpositions of all the basis states which include 
a single qubit in the 1 state and all the other qubits
in the zero state. It takes the form

\begin{equation}
|\psi_{\rm W}\rangle = \frac{1}{\sqrt n} \sum_{k=1}^{n} |2^{k-1} \rangle. 
\end{equation}

\noindent
In the case of two qubits it coincides with the Bell state
$|\psi_{+}\rangle = \frac{1}{\sqrt{2}}(|01\rangle+|10\rangle)$,
while in the case of three qubits it takes the form
$\frac{1}{\sqrt{3}}(|001\rangle+|010\rangle+|100\rangle)$. 
The overlap function for these states take the form

\begin{equation}
P(\theta_1,\dots,\theta_n) 
= \frac{1}{n} \left( 
\sum_{k=1}^n \sin \theta_k \prod_{k^{\prime} \ne k} \cos \theta_{k^{\prime}}
\right)^2.
\end{equation}

\noindent
Taking derivatives with respect to the $\theta_k$'s 
we obtain that the maximal value is obtained at
$\sin\theta_k = 1/\sqrt n$,
$\cos\theta_k = \sqrt{1-1/n}$,
$k=1,\dots,n$.
The maximal value is
 
\begin{equation}
P_{\max}(\psi_{\rm W}) = \left(1 - \frac{1}{n} \right)^{n-1},
\label{eq:wclassfin}
\end{equation}

\noindent
which converges to $1/e$ as the number of qubits increases. 
The Groverian measure then converges to
$G(\psi_{\rm W})=\sqrt{1-1/e}$ 
at 
$n \rightarrow \infty$, 
which is higher than the value for the GHZ states
but lower than $1$. 
We will now examine a class of 
strongly entangled states,
in which the Groverian measure converges to 1 at 
$n \rightarrow \infty$. 

\subsubsection{The balanced states}

Consider a state of an even number, $n$, of qubits, that consists of an equal
superposition of all the balanced basis states, namely of all those
basis states in which the number of 0's is equal to the number of 1's.
This state takes the form

\begin{equation}
| \psi \rangle = 
{\frac{1}{\sqrt{K}}}
(|0\dots01\dots1\rangle + {\rm Permutations}).
\end{equation}

\noindent
where the binomial coefficient

\begin{equation}
K =
\left( 
\begin{array}{c} 
n \\ 
{ n/2 } 
\end{array} 
\right)
\end{equation}

\noindent
is equal to the number of different  
permutations of $n$ bits, where
$n/2$ of them are 
in the 0 state and the other $n/2$ are in the 1 state.
The overlap function for this state is

\begin{equation}
P(\theta_1,\dots,\theta_n,\psi) = 
\sqrt{K}
(\cos \theta_1 
\cdots 
\cos \theta_{n/2} 
\sin \theta_{n/2+1} 
\cdots
\sin \theta_{n} 
+ {\rm Permutations}).
\end{equation}

\noindent
Taking the derivatives and solving for
$\partial P(\theta_1,\dots,\theta_n,\psi) / \partial \theta_k = 0$,
we obtain that the maximum is at
$\theta_k=\pi/4$, $k=1,\dots,n$,
and 

\begin{equation}
P_{\max}(\psi) = K \left(\frac{1}{2}\right)^{n} 
\end{equation}

\noindent
In the limit of $n \rightarrow \infty$
one can use the Stirling approximation,
that gives rise to

\begin{equation}
P_{\max}(\psi) \rightarrow \sqrt{\frac{2}{\pi}} \frac{1}{\sqrt{n}}.
\end{equation}

\noindent
Since $P_{\rm max}(\psi)$ converges to zero, the Groverian measure
$G(\psi)$
of the balanced states converges to 1
as $n \rightarrow \infty$.

\section{Numerical calculations of the Groverian measure}

\subsection{The numerical procedure}

In order to evaluate the Groverian entanglement measure of a given
pure state
$| \psi \rangle$
of $n$ qubits
one has to find the
maximal overlap of this state with any product state with the
same number of qubits.
The overlap is given by the function
$P(\theta_1,\dots,\theta_n,\varphi_1,\dots,\varphi_n,\psi)$.
This is thus a maximization problem in a 
$2n$-dimensional space 
(which is reduced to $n$ dimensions if all the amplitudes are real).
In general, this problem cannot be solved analytically, and therefore
the use of numerical calculations is essential.

The vector space in which the maximization is performed is 

\begin{equation}\label{eq:compVector}
\vec{r} = \left(\theta_1,\dots,\theta_n,\varphi_1,\dots,\varphi_n \right)
\end{equation}

\noindent
for the general case in which the amplitudes are complex
[in the special case of real amplitudes it is reduced to
$\vec{r} = \left(\theta_1,\ldots,\theta_n\right)$].
The optimization is done using the steepest descend method, namely

\begin{equation}
\frac{d\vec{r}}{dt} = C \ \vec{\nabla} 
P(\theta_1,\dots,\theta_n,\varphi_1,\dots,\varphi_n,\psi),
\end{equation}

\noindent
where $C>0$ is a constant and
$\vec{\nabla} = (\partial/\partial \theta_1,\dots,\partial/\partial\theta_n,
\partial/\partial \varphi_1,\dots,\partial/\partial \varphi_n)$.

\noindent
For a given initial point, $\vec r$, the steepest descent method converge
to a nearby local maximum. In order to obtain the global maximum,
the calculation is repeated with a large number of random initial points.
The largest value among all the local maxima that are reached is then
picked as the numerical result for $P_{\rm max}$.

In general, the function 
$P(\theta_1,\dots,\theta_n,\varphi_1,\dots,\varphi_n,\psi)$
consists of a sum of products of trigonometric functions.
Each product includes $n$ sine or cosine functions and 
is thus periodic in all directions.
The number of terms in the sum increases exponentially with $n$,
making the maximization problem more difficult.
We find that the number of local minima also increases with $n$.
Thus, as $n$ increases one needs more calls to the steepest descent
program with random initial points. 

To exemplify the effect of the maximization process we consider the
family of states given by

\begin{equation}
| \psi \rangle = 
a_{\eta} |\eta\rangle + 
a_{\rm GHZ } |\psi_{\rm GHZ} \rangle,
\label{eq:etaghz}
\end{equation}

\noindent
where $0 \le a_{\rm GHZ} \le 1$
and since the two states are not orthogonal we obtain from the normalization condition
that

\begin{equation}
a_{\eta} = -\sqrt{{2}/{N}}\ a_{\rm GHZ} + \sqrt{1-(1-2/N) a^2_{\rm GHZ} }. 
\label{eq:aetaexp}
\end{equation}

\noindent
In Fig. 1 we present the success probability
$P_{\rm s}(\psi)$ 
(dashed line)
of Grover's search with the initial state $|\psi\rangle$
as well as 
$P_{\rm max}(\psi)$
(solid line)
obtained from the numerical procedure,
as a function of 
${a^2_{\rm GHZ}}$, 
for 12 qubits.
The two functions follow the same path, decreasing linearly
as
${a^2_{\rm GHZ}}$
increases,
up to 
${a^2_{\rm GHZ}}\simeq 0.65$.
This means that for the states in this range the success probability
cannot be increased by local unitary operations.
As the two functions depart,
$P_{\rm max}(\psi)$
starts to increase while
$P_{\rm s}(\psi)$ 
continues to decrease.
The size of the gap between the two functions represents
the effect of the maximization process.

In Fig. 2 we show the Groverian measure of the generalized GHZ states
[Eq. (\ref{eq:Genghz})]
of 12 qubits as a function of $|a_0|^2$. The numerical
results ($\circ$) coincide with the analytical results (solid line).
The graph is symmetric around
$|a_0|^2=1/2$,
where the largest value of $G(\psi)$ is obtained. 
At this point, for real and positive amplitudes the state
$|\psi_{\rm GHZ}\rangle$ 
is obtained.

\subsection{Entanglement during the operation of Grover's algorithm:}

There are indications 
that entanglement plays an important role 
in making quantum algorithms more efficient than their classical counterparts.
Therefore, it is interesting to see how entanglement is generated 
during the
operation of quantum algorithms.
Grover's algorithm is particularly suitable for this study because it
consists of a large number of iterations of the same set of operations.
Furthermore, for any given initial state 
$|\psi(0)\rangle$, 
the amplitudes of the state
$|\psi(t)\rangle$, 
obtained after $t$ Grover iterations, can be calculated analytically
\cite{Biham1999,Biham2003}.

In Fig. 3 we present the Groverian entanglement measure of the 
states 
$|\psi(t)\rangle$, 
obtained after  
$t=0,1,\dots,50$
iterations of Grover's algorithm with
$n=12$ qubits for 
the initial state is 
$|\psi(0)\rangle = | \eta \rangle$
(dashed line).
It increases with the time until it
reaches its highest value of $1/\sqrt{2}$ after 
$25$ iterations.
Then it follows the same path downwards reaching zero value
after $50$ iterations, where the marked state is reached. 
The case in which the initial state is entangled is also
considered. To this end we construct the state

\begin{equation}
| \psi \rangle = 
a_{\rm even} | \psi_{\rm even} \rangle 
+
a_{\rm odd} | \psi_{\rm odd} \rangle
\label{eq:evenodd} 
\end{equation}

\noindent
where 
$| \psi_{\rm even} \rangle$
is a normalized superposition of all the
$N/2$ basis states that include an even number of
1's,
while 
$| \psi_{\rm odd} \rangle$
is a normalized superposition of all the other basis
states, that include an odd number of 1's.
Apart from the initial state $| \eta \rangle$
that is obtained for
$a_{\rm even}=1/\sqrt{2}$,
we also show the results for 
$a_{\rm even} = 0.984$
(dotted line),
$0.994$ (dashed-dotted line)
and 1 (solid line).
The last state in this list is the one in which
the amplitudes of
all the basis states that have an odd number
of 1's vanish.
This state can be obtained from the GHZ state by applying the
Hadamard transform on all the qubits.

In Fig. 4 we show $G(\psi)$
for the states 
$|\psi(t)\rangle$
obtained after
$t$ Grover iterations
where
$|\psi(0)\rangle = | \eta \rangle$
for 12 qubits and two marked states.
In case that the marked states are
$|0\rangle$
and
$|N-1\rangle$
(solid line)
the resulting state
after $\tau$ iterations is
$|\psi_{\rm GHZ}\rangle$.
In case that the two marked states
are 
$|0\rangle$
and 
$|1\rangle$
(dashed line)
the resulting state is 
a superposition of these two states,
which is
not entangled.
Interestingly, during the first $\tau/2$ iterations
these two functions nearly coincide.

In Fig. 5 we present
the Groverian measure of the states
$|\psi(t)\rangle$
obtained after $t$ Grover iterations,
where
$|\psi(0)\rangle = | \eta \rangle$
for 12 qubits and 12 marked states.
When the
set of marked states 
consists of 
$|00\dots01\rangle$,
$|00\dots10\rangle$, 
$\dots$,
$|10\dots0\rangle$,
namely the basis states of non-zero amplitudes
in the W state (solid line),
the register approaches the W state
after 14 iterations. 
When the marked states are
$|i\rangle$, $i=0,\dots,11$
(dashed line)
the resulting state 
is entangled, but exhibits
a smaller value of 
$G(\psi)$.

\subsection{Random states of $n$ qubits}

How entangled is a randomly chosen pure quantum state
of $n$ qubits? 
To study this question we pick random pure states
and evaluate their Groverian entanglement measure using
the numerical procedure.

Consider a random pure state 

\begin{equation}
|\psi\rangle = \sum_{i=0}^{N-1} a_i |i\rangle 
\end{equation}

\noindent
of $n$ qubits,
where $a_i = |a_i| e^{i \alpha_i}$.
Such random states can be obtained 
as follows. In the first step one 
draws $N$ numbers,
$a_i$, $i=0,\dots,N-1$,
independently
from a Gaussian distribution
centered at 0, with a standard deviation
$\sigma=1$.
These numbers are then normalized according to

\begin{equation}
a_i \rightarrow \frac{a_i}{\sqrt{\sum_{j=0}^{N-1} |a_j|^2}}.
\end{equation}

\noindent
The moduli
$|a_i|$, $i=0,\dots,N-1$
are then obtained as the absolute values
of the $a_i$'s.
The arguments
$\alpha_i$, $i=0,\dots,N-1$
are drawn 
from a homogeneous distribution
in the range $[0,2\pi)$.
Due to the properties of the Gaussian distribution,
the resulting states are distributed randomly and 
isotropically in the Hilbert space
that consists of all the pure states of $n$ qubits.

From the normalization condition we obtain that
the second moment of the distribution of the moduli
of the amplitudes,

\begin{equation}
\overline{|a|^2} = {\frac{1}{N}} \sum_{i=0}^{N-1} |a_i|^2,
\end{equation}

\noindent
satisfies
$\overline{|a|^2} = 1/N$,
namely,
its square root is of order
$1/\sqrt{N}$.
In the case of the randomly chosen states,
due to the random phases,
the average amplitude 
$\bar{a}$
[given by Eq. (\ref{eq:abar})]
will be of the order of
$\bar{a} = 1/N$.
Therefore, the probability of success 
$P_{\rm s}(\psi)$
of Grover's algorithm using
a random state as the initial state will vanish like
$1/N$.
Our analysis of these states using the numerical procedure
shows that for the random states the effect of the optimization procedure
is negligible.
As a result, $P_{\rm max}(\psi)$
also vanishes like
$1/N$. 
Thus, the entanglement measure
$G(\psi) \rightarrow 1$
as the number of qubits increases
indicating that the vast majority of the states
of multiple qubits are very highly entangled.

\section{Summary and discussion} 
 
The Groverian entanglement measure was applied to 
characterize a variety of pure quantum states of multiple qubits.
For certain classes of states of high symmetry,
the Groverian measure was calculated analytically.
In order to evaluate it for
arbitrary states with complex amplitudes, 
a numerical minimization procedure, based on the 
steepest descent algorithm was developed.
It was used in order to evaluate the amount of entanglement
generated by Grover's algorithm 
for different initial states and for different sets of 
marked states.
It was also shown that the typical pure states 
of $n$ qubits
obtained by
random sampling 
are highly entangled, namely for these
states
$G(\psi)=1$ up to corrections of order $1/N$.

In recent years several entanglement monotones were proposed
as measures of entanglement of multiple qubits
\cite{Vedral1997,Vedral1998,Barnum2001,Meyer2002,Leifer2003,Wei2003}.
Unlike the case of pure states of two qubits, in which
the von Neumann entropy provides a complete characterization
of the entanglement, multiple qubit states support a large 
number of different measures. 
As a result there is no meaningful way to compare
between the different measures.
It seems that the issue of what measure is relevant 
depends on the specific physical or operational context.
The actual evaluation of entanglement measures 
turns out to be a difficult computational problem.
This is due to the fact that these measures are typically
defined as an extremum of some multi-variable function.
A singular result in this context is the explicit
formula for the entanglement of formation of mixed states
of two qubits, obtained in Refs.
\cite{Hill1997,Wootters1998}.

The Groverian measure was originally introduced for the case
of a single marked state. Recently, it was shown that the 
same result for 
$P_{\rm max}(\psi)$
is obtained in the case of $r$ marked states,
up to a correction of order $r/N$
\cite{Biham2003},
thus removing the restriction of a single marked state. 
Grover's algorithm can be generalized by replacing the
Hadamard transform by an arbitrary unitary operator $U$
on the $n$ qubits
\cite{Grover1998,Long1999,Gingrich2000,Biham2001}.
Using the generalized algorithm to evaluate
$P_{\rm max}(\psi)$,
one observes that the same result is obtained as
long as the operator $U$ is a tensor product of
$n$ unitary single qubit operators.
In case that the operator $U$ creates entanglement
between the qubits in the register it cannot be
used to evaluate the Groverian measure.

Interestingly, the Groverian measure coincides for pure states
with the measure proposed by
Vedral, Plenio, Rippin and Knight
\cite{Vedral1997,Vedral1998}.
While that measure also applies in the case of mixed states
of multiple qubits, we have not been able to extend the
operational interpretation of the Groverian measure
beyond the case of pure states
\cite{Biham2002}.
It would be interesting to examine the relevance of the
Groverian measure to other quantum algorithms such as
Shor's algorithm. 
Further studies of the Groverian measure and related
concepts will hopefully contribute to the understanding
of the role of entanglement in making quantum algorithms
powerful.

\acknowledgments
This work was supported by the EU Grant No. IST-1999-11234.

\newpage
\clearpage

\begin{figure}
\caption{The success probability $P_{\rm s}(\psi)$
of Grover's algorithm using $|\psi\rangle$
as the initial state (dashed line), 
and the maximal success probability
$P_{\rm max}(\psi)$ 
(solid line),
for
$|\psi\rangle = a_{\eta} |\eta\rangle + a_{\rm GHZ} |\psi_{\rm GHZ}\rangle$,
as a function of 
$a^2_{\rm GHZ}$.
The two functions coincide up to 
$a^2_{\rm GHZ}\simeq 0.65$,
while above this point a gap appears.
The gap broadens
as the GHZ state
is approached, demonstrating the effect of the 
maximization by the
local pre-processing.
}
\label{fig:PsPmax}
\end{figure}

\begin{figure}
\caption{Analytical results (solid line) and numerical 
results ($\circ$) for
the Groverian entanglement measure of generalized GHZ states 
[Eq.~(\ref{eq:Genghz})] 
as a function of $|a_0|^2$. 
The largest value of $G(\psi)=1/\sqrt{2}$ 
is obtained for 
$|a_0|^2=1/2$,
where the state coincides with the GHZ state.
}
\label{fig:ghz}
\end{figure}

\begin{figure}
\caption{The Groverian entanglement measure $G(\psi(t))$
for states that are generated by 
Grover's algorithm as a function of the number of iterations
$t$ for $n=12$ and one marked state,
where the initial state is
$|\psi(0)\rangle$.
The curves were obtained for the states 
$|\psi(0)\rangle$
given by Eq.
(\ref{eq:evenodd})
with
$a_{\rm even} =1/\sqrt{2}$, namely the $|\eta\rangle$ state
(dashed line),
0.984 (dotted line), 0.994 (dashed-dotted line) and 1,
namely the state $H^{\otimes n} |\psi_{\rm GHZ}\rangle$ 
(solid line).
}
\label{fig:grover}
\end{figure}

\begin{figure}
\caption{
The Groverian measure
for the states 
$|\psi(t)\rangle$
obtained after
$t$ Grover iterations
where
$|\psi(0)\rangle = | \eta \rangle$,
for 12 qubits and two marked states.
In case that
$|0\rangle$
and
$|N-1\rangle$
are marked
(solid line)
the resulting state
after $\tau$ iterations is
$|\psi_{\rm GHZ}\rangle$.
If, instead, the marked states are
$|0\rangle$
and 
$|1\rangle$
(dashed line),
the register approaches
a superposition of these two states,
which is not entangled.
}
\label{fig:etaghz}
\end{figure}
 
\begin{figure}
\caption{
The measure
$G(\psi(t))$
for states that are generated by 
Grover's algorithm as a function of the number of iterations
$t$ for $n=12$ and 12 marked states,
where the initial state is
$|\eta\rangle$.
When
the marked states are 
$|00\dots01\rangle$,
$|00\dots10\rangle$, 
$\dots$,
$|10\dots0\rangle$
(solid line),
the state obtained after 14 iterations is the W state
(up to a tiny correction due to the discrete nature of
the iterations).
When the marked states are
$|i\rangle$, $i=0,\dots,11$
(dashed line),
the resulting state is not as
strongly entangled. 
}
\label{fig:etaw}
\end{figure}

\newpage
\clearpage
\includegraphics[width=17cm]{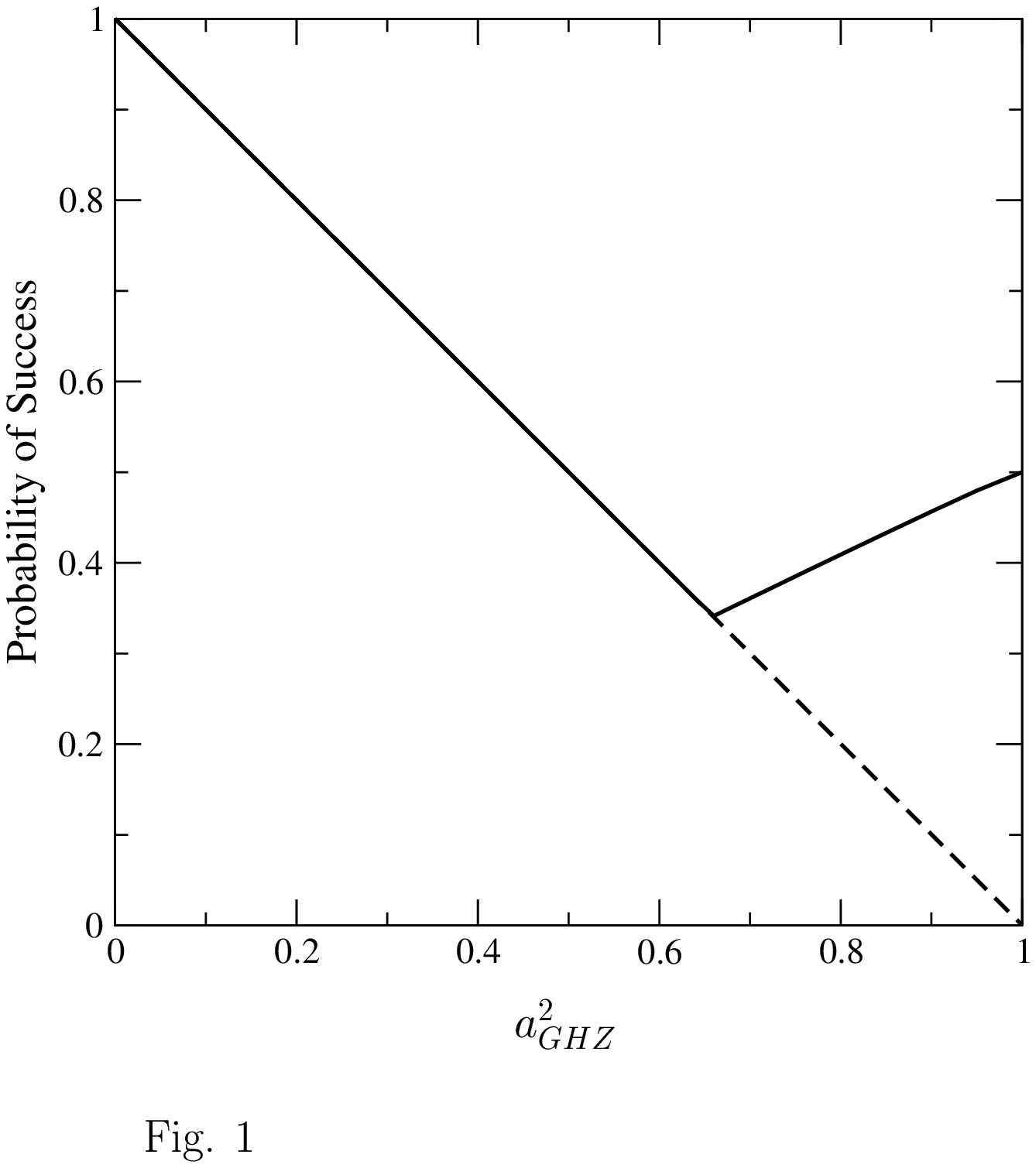}
\newpage
\includegraphics[width=17cm]{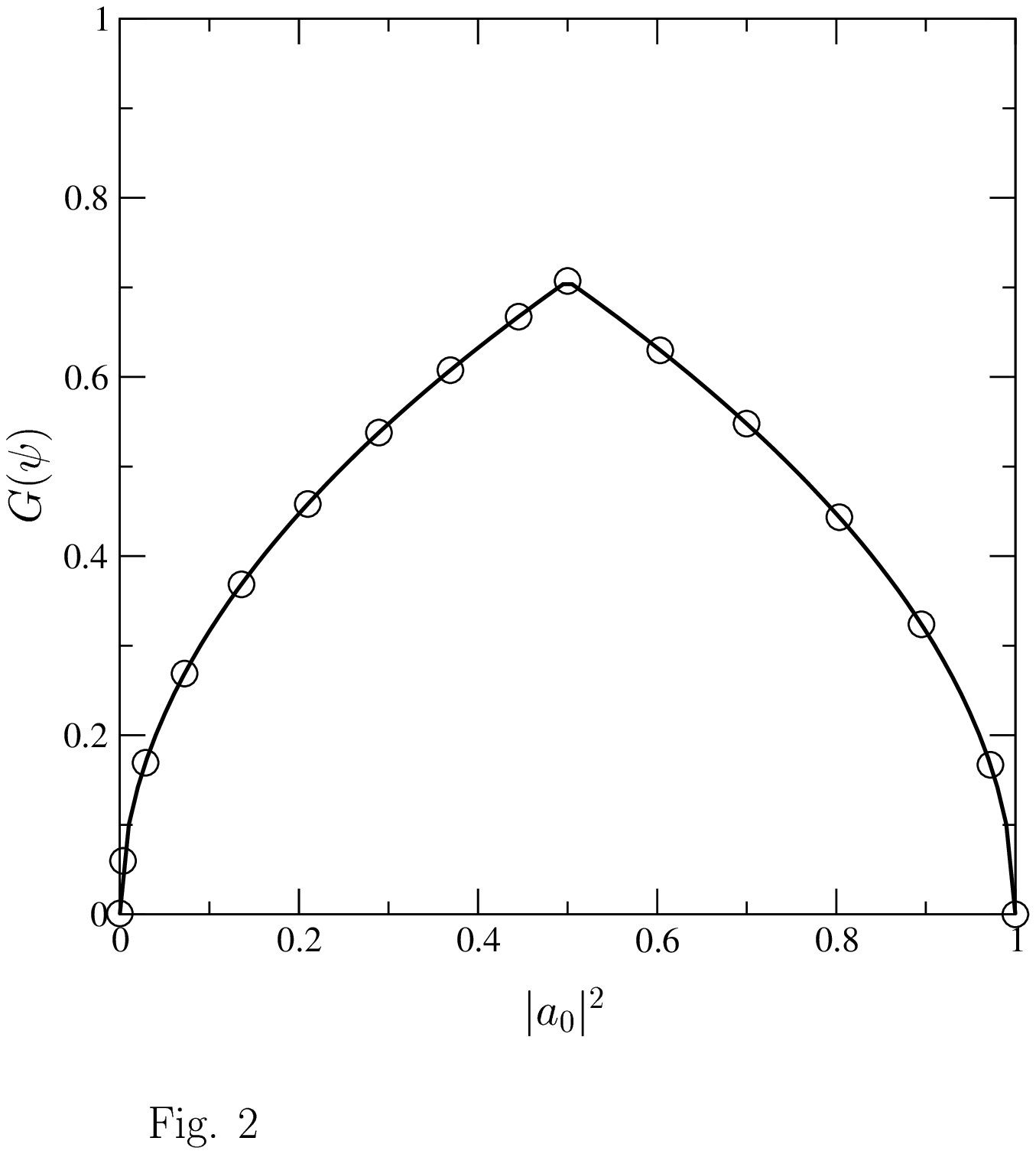}
\newpage
\includegraphics[width=17cm]{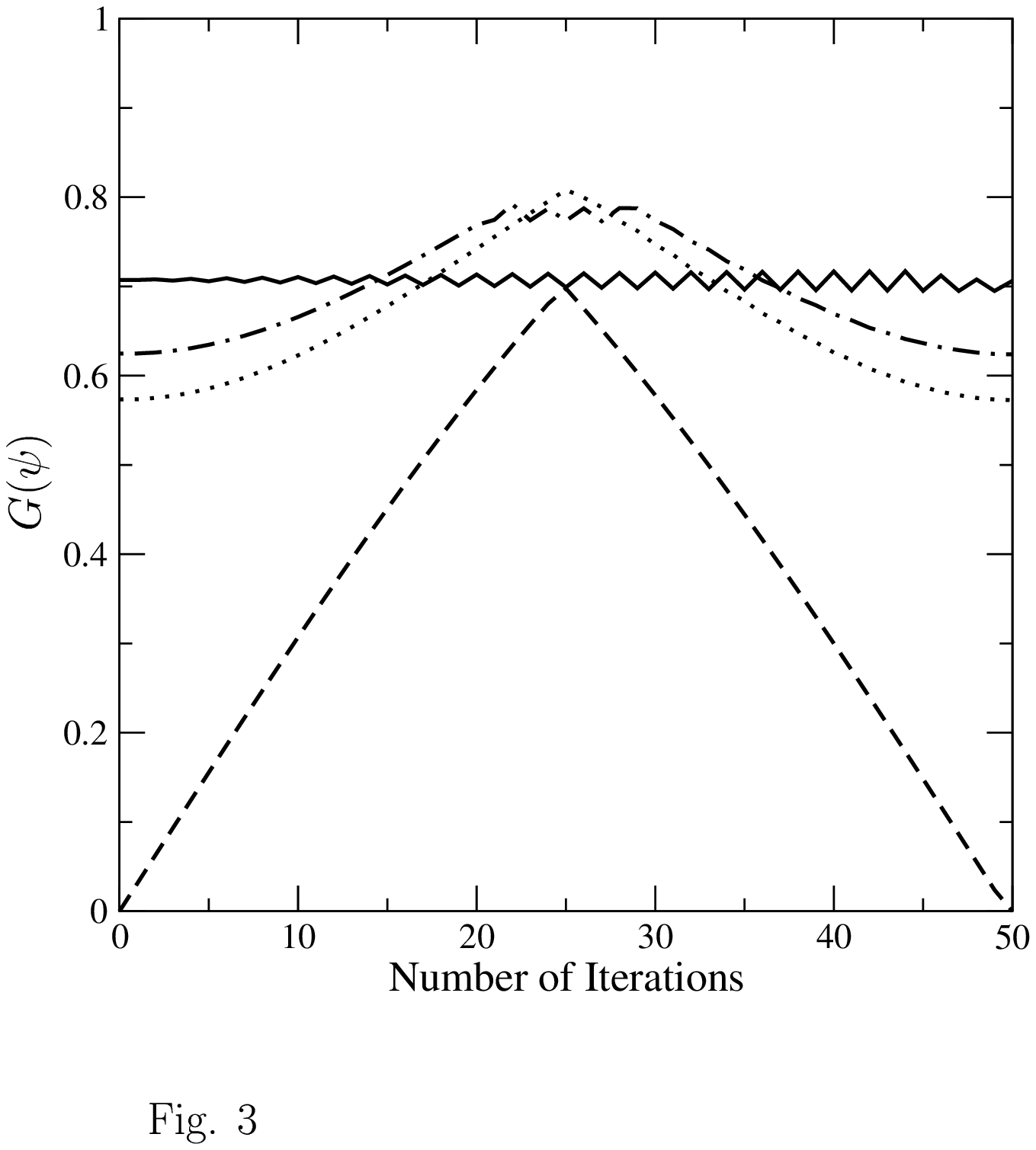}
\newpage
\includegraphics[width=17cm]{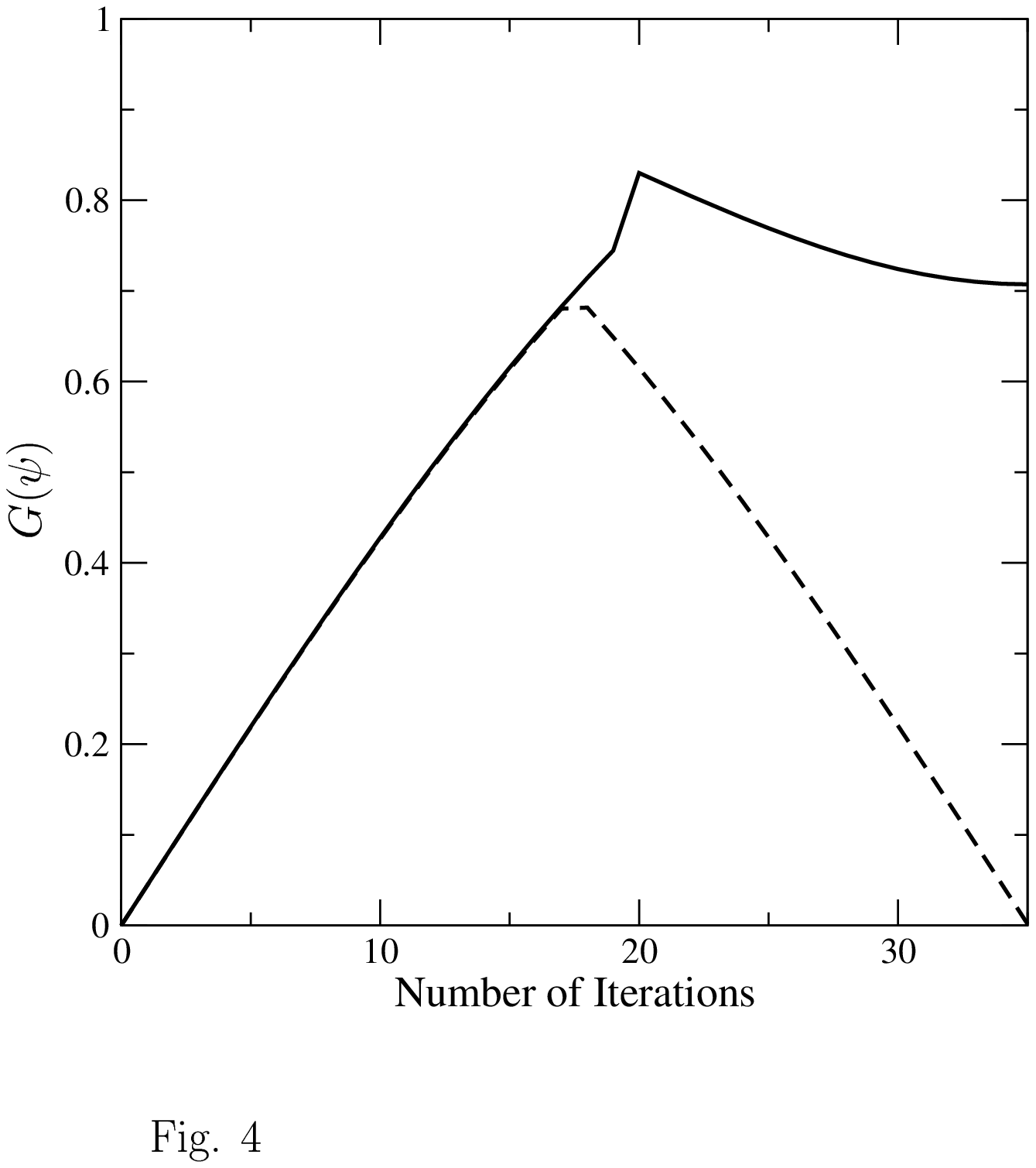}
\newpage
\includegraphics[width=17cm]{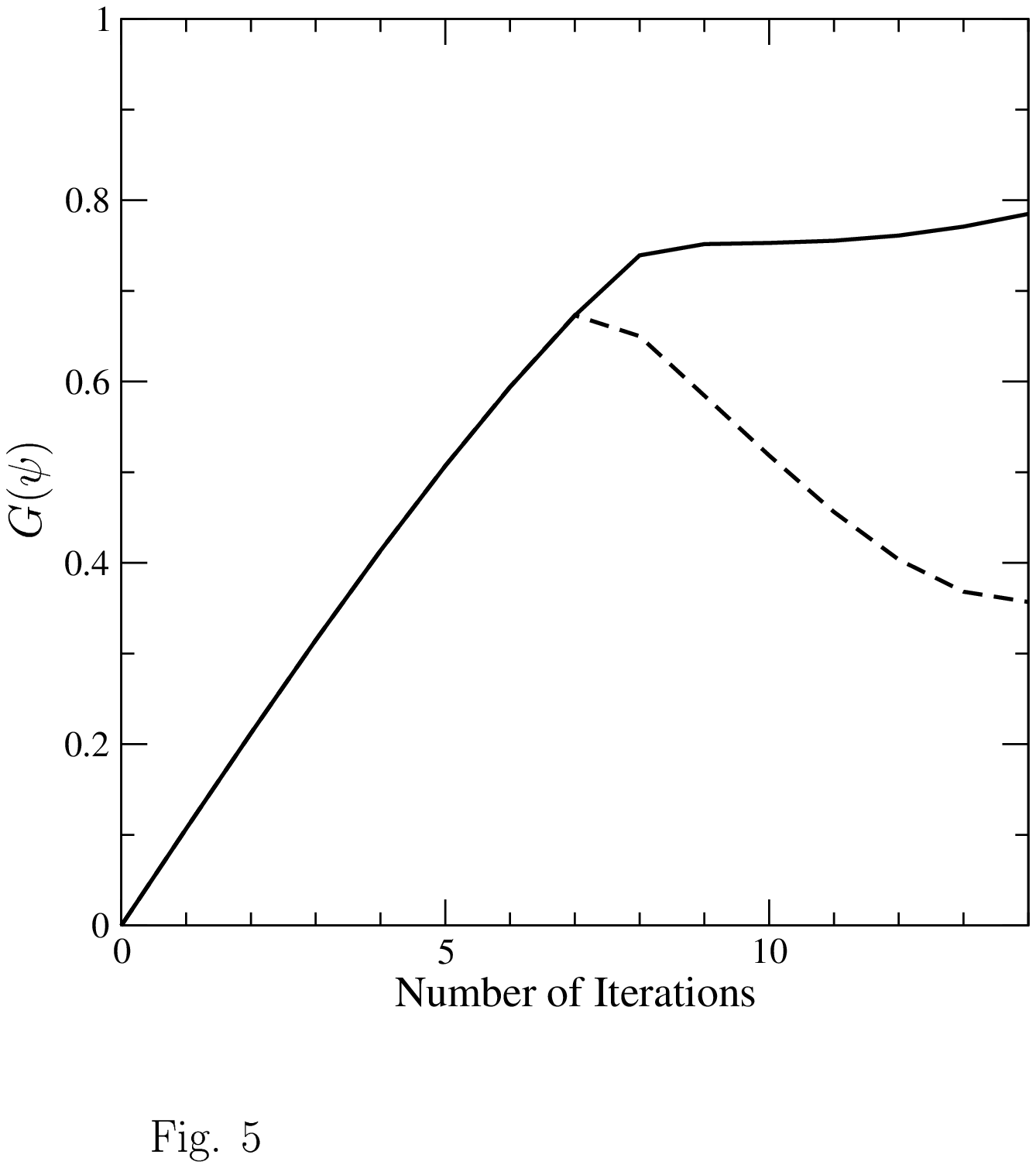}
\end{document}